\documentclass{article}
\usepackage{amssymb}
\usepackage{color}
\usepackage{hyperref}
\usepackage{amsmath}

\numberwithin{equation}{section}

\newcommand{\p}[1]{(\ref{#1})}

\newcommand{\cF}{{\cal F}}

\newcommand{\cB}{{\cal B}}

\newcommand{\cA}{{\cal A}}
\newcommand{\cbA}{\overline{\cal A}}
\newcommand{\cW}{{\cal W}{}\,}
\newcommand{\cbW}{\overline{\cal W}{}\,}

\newcommand{\bX}{{\overline X}{}}

\newcommand{\bD}{{\overline D}{}}

\newcommand{\bx}{{\bar x}}

\newcommand{\be}{\begin{equation}}
\newcommand{\ee}{\end{equation}}
\newcommand{\bea}{\begin{eqnarray}}
\newcommand{\eea}{\end{eqnarray}}

\newcommand{\ba}{\begin{array}} \newcommand{\ea}{\end{array}}
\newcommand{\ds}{\displaystyle}

\begin{document}
\thispagestyle{empty}
\vspace{2cm}
\begin{flushright}
%Semi-final version \\
%Version 2.1 \\
%\today\\
\end{flushright}\vspace{2cm}
\begin{center}
{\Large\bf On the road to $N=2$ supersymmetric Born-Infeld action}
\end{center}
\vspace{1cm}

\begin{center}
{\large\bf S.~Bellucci${}^a$, S.~Krivonos${}^{b}$, A.~Shcherbakov${}^a$ and
A.~Sutulin${}^{b}$ }
\end{center}

\begin{center}
${}^a$ {\it
INFN-Laboratori Nazionali di Frascati,
Via E. Fermi 40, 00044 Frascati, Italy} \vspace{0.2cm}

${}^b$ {\it
Bogoliubov  Laboratory of Theoretical Physics, JINR,
141980 Dubna, Russia} \vspace{1.2cm}

{\tt bellucci,ashcherb@lnf.infn.it, krivonos,sutulin@theor.jinr.ru}

\end{center}
\vspace{2cm}

\begin{abstract}\noindent
We analyze the exact perturbative solution of~$N=2$ Born-Infeld theory which is believed to be defined by Ketov's equation.  This equation can be considered as a truncation of an infinite system of coupled differential equations defining Born-Infeld action with one manifest~$N=2$ and one hidden~$N=2$ supersymmetries. We explicitly demonstrate that infinitely many new structures appear in the higher orders of the perturbative solution to Ketov's equation. Thus, the full solution cannot be represented as a function depending on {\it a finite number} of its arguments. We propose a mechanism for generating the new structures in the solution and show how it works up to~18-th order. Finally, we discuss two new superfield actions containing an infinite number of terms and sharing some common features with~$N=2$ supersymmetric Born-Infeld  action.

\end{abstract}

\newpage
\setcounter{page}{1}
\section{Introduction}
It is known from papers~\cite{BG1,RT} that~$N=1$ supersymmetric Born-Infeld (BI) action \cite{CF} could be uniquely constructed if one imposes that  the theory possesses a global~$N=2$ supersymmetry, half of which is manifest and the other half is spontaneously broken.
The corresponding Goldstone superfield is a~$N=1$ vector, or Maxwell, supermultiplet~$W_\alpha$. The~$N=1$ BI action possesses~$U(1)$ duality~\cite{KT1,KT2} and has a remarkable compact form~\cite{BG1,RT}. The corresponding superspace chiral Lagrangian density~$A$ was argued to be a solution to the following equation:
$$\ds A = \frac{W^2}{1-\frac14 \bar D^2 \bar A}.$$

Trying to construct a $N=2$ supersymmetric BI action raises much richer and more difficult issues. This is mainly the reason why the complete form of this action is still missing. There exist several approaches to the problem of constructing the~$N=2$ BI action.

One of them is a generalization of the above mentioned approach proposed in papers~\cite{ketov1,ketov2}. It consists in imposing a constraint
\be\label{eq-1}
\cA=\cW^2 +\frac{1}{2} \cA \bD^4 \cbA
\ee
(which will be called as Ketov equation within this paper) on the superspace chiral Lagrangian density~$\cA$ which defines the~$N=2$ supersymmetric action of BI theory
\be\label{action-1}
S = \int d^4x d^4\theta \; \cA + \int d^4 x d^4 {\bar\theta}\; \cbA
\ee
in terms of an off-shell~$N=2$ vector superfield~$\cW$ which comprises the Maxwell field strength among its components.
In~\cite{ketov1,ketov2} it was claimed that with respect to the electric-magnetic duality the action~\p{action-1} is self-dual, while the existence of hidden $N=2$ supersymmetry has been just hypothesized. There it was also asserted that  Eq.~\p{action-1} can be written in the following form:
\be\label{action0}
S = \int d^4 x d^4 \theta \; \cW^2 +\int d^4 x d^4 \bar\theta\;  \cbW^2 +2
\int d^4 x d^8 \theta \; \Upsilon( x, \bx) \cW^2 \cbW^2,
\ee
where~$\Upsilon( x, \bx)$ is a function depending only on the following arguments:
\be\label{var0}
x = D^4 \cW^2, \qquad \bx= \bD^4 \cbW^2.
\ee
Despite its simplicity, eq.~\p{eq-1} is hard to solve and its full solution is yet to be found. The solution can be found iteratively by performing a perturbative expansion in~$\cW$ and the lower approximations evidently supports the assertion~\p{action0}.

In papers~\cite{KT1,KT2} it was proved that the action~\p{action-1} with Lagrangian $\cA$ defined by eq.~\p{eq-1}
indeed describes a self-dual theory, and it was shown that in fact the iterative solution to the constraint~\p{eq-1} cannot be represented by a function~$\Upsilon( x, \bx)$ depending on~$x$ and~$\bx$ only. This fact shows up for the first time at the 10-th order in the $\cW$ expansion. %This funny situation with a good superfield constraint~\p{eq-1} and its strange  ``solution'' exists not very long.

Another approach was developed in a series of papers~\cite{BIK1,BIK2,BIK3} where the construction of~$N=2$ BI action was proposed assuming the presence of a hidden~$N=2$ supersymmetry being realized nonlinearly. In this approach, the superspace chiral Langrangian density is a solution to a system of infinitely many coupled differential equations.  It is noteworthy that eq.~\p{eq-1} is just a truncation of this system of equations. The truncation  consists in dropping out terms with higher order differential operators so that one of the equations of this system decouples from the rest.

This infinite system of equations can also be solved by iterations up to any desired order in the fields, but the solution contains  terms with the explicit presence of d'Alembert operator.

Last but not least, an approach was developed in~\cite{KT1}. It is based on self-duality and invariance under nonlinear shifts of the vector superfield~$\cW$; the corresponding action coincides with the action of the approach~\cite{BIK1,BIK2,BIK3}.

In the present paper we investigate the existence of a compact form of the general solution of equation~\p{eq-1}.
In section~\ref{prem} we analyze higher order terms in the perturbative expansion of the solution and explicitly demonstrate
that each order gives rise to new structures. This fact makes problematic the representation of the full solution
as some function depending on {\it a finite number} of its arguments. In section~\ref{KEq} we perform the analysis of a
``linear'' approximation to Ketov equation and advocate the idea of introducing differential operators instead of the
variables~$x$ and~$\bx$ in~\p{var0}. The corresponding action acquires the form~\p{action0} but with the pseudo-differential
operator which replaces the function~$\Upsilon$. Passing from functions to operators helps us to construct the action,
which coincides with the exact solution to equation~\p{eq-1} up to~10-th order. The next orders call for new structures.
It turns out that with one additional operator one could succeed in getting up to 20-th order, where the introduction of new
operators becomes unavoidable.

The existence (at least partially) of the action written as pseudo-differential operators acting on some superfields and,
therefore, containing an infinite number of terms leads to the question about the existence of another more appropriate framework for constructing~$N=2$ BI action.

\section{Preliminaries}\label{prem}
In papers~\cite{BIK1,BIK2,BIK3} it is argued that the basic object of the hypothetical~$N=2$ BI theory should be a complex chiral scalar~$N=2$ superfield strength~$\cW$ subjected to the chirality condition~(a) and the Bianchi identity~(b)
\be\label{W}
(a)\quad \bD_{{\dot\alpha}i}\cW=0,\quad D^i_\alpha \cbW =0, \qquad (b) \quad D^{ik}\cW=\bD^{ik}\cbW,
\ee
where
$$D^{ij}\equiv D^{\alpha i}D_\alpha^j, \qquad \bD^{ij} \equiv \bD^i_{\dot\alpha} \bD^{\dot\alpha j} $$
and the covariant derivatives satisfy the following algebra
\be\label{def}
\left\{ D^i_\alpha, \bD_{\dot\alpha j}\right\}=-2 i \delta^i_j \partial_{\alpha \dot\alpha}.
\ee
In~\cite{BIK1} it has been proved that the hidden~$N=2$ supersymmetry (together with central charge transformations)
can be realized on the basic object~$\cW$ as
\be\label{SUSY2}
\delta \cW=f\left( 1- \frac{1}{2} \bD^4 \cA_0\right) +\frac{1}{4}{\bar f} \Box \cA_0-
\frac{i}{4} \bD^{i\dot\alpha}{\bar f}\; D^\alpha_i \partial_{\alpha\dot\alpha}\cA_0,
%\quad\delta\cbW =\left(\delta \cW\right)^\star ,
\ee
where
$$%\be\label{def1}
D^4 \equiv \frac{1}{48} D^{\alpha i}D_\alpha^j D_i^\beta D_{\beta j}, \qquad
\bD^4 \equiv \frac{1}{48} \bD^i_{\dot\alpha}\bD^{\dot\alpha j}\bD_{\dot\beta i}\bD_j^{\dot\beta},\qquad
\Box\equiv \partial_{\alpha\dot\alpha}\partial^{\alpha\dot\alpha}.
$$%\ee
The complex function~$f$ is defined as follows:
$$%\be
f=c +2 i \eta^{i \alpha}\theta_{i\alpha}, \quad {\bar f}={\bar c}- 2i {\bar\eta}{}^i_{\dot\alpha}
{\bar\theta}{}^{\dot\alpha}_i,
$$%\ee
and it collects parameters of the second~$N=2$ supersymmetry~$\eta^{i\alpha}$,~$\bar\eta{}^i_{\dot\alpha}$ and those of the
central charge transformations~$c$,~${\bar c}$. The complex chiral function~$\cA_0$ is a good candidate for~$N=2$ Lagrangian density, because the action
\be\label{action1}
S = \int d^4x d^4\theta \; \cA_0 + \int d^4 x d^4 {\bar\theta}\; \cbA_0
\ee
is invariant with respect to the hidden~$N=2$ supersymmetry~\cite{BIK1}. The only problem is to express~$\cA_0$
in terms of~$\cW$. This step is the most complicated part of the construction, because the function~$\cA_0$ obeys
the highly non-trivial equation \cite{BIK1}
\be\label{eq1}
\cA_0=\cW^2 +\frac{1}{2} \cA_0 \bD^4 \cbA_0+\frac{1}{2}\sum_{k=1} \frac{(-1)^k}{8^k}\cA_k \Box^k \bD^4 \cbA_k.
\ee
The chiral functions~$\cA_k$ entering~\p{eq1}, in turn, satisfy nonlinear equations. These equations form an infinite set of entangled differential equations. This system of equations cannot be reduced to a finite subsystem~-- the number of the functions~$\cA_k$ is fundamentally infinite.
Until now only lower order terms (up to 10-th) are known in the perturbative expansion  of~$\cA_0$ over~$\cW$.

Equation~\p{eq1} (as well as the accompanying ones for $\cA_k$ \cite{BIK1,BIK2,BIK3}) looks so complicated that it seems hardly possible to construct its full solution. One may try to simplify the situation dropping down terms containing higher orders of differential operators, i.e. ignoring~$\Box$-dependent terms in~\p{eq1}. Doing so we will deal with the following equation defining the Lagrangian density in the action \p{action1}:
\be\label{eq2}
\cA_0=\cW^2 +\frac{1}{2} \cA_0 \bD^4 \cbA_0.
\ee
Clearly, with such a truncation we lost the hidden~$N=2$ supersymmetry, but some important properties of the full~$N=2$ BI
theory would survive. Namely, the equation~\p{eq2} is just the Ketov's equation~\p{eq-1}. Again, the solution to equation~\p{eq2} is
known as a series expansion over~$\cW$ up to few orders. In what follows we analyze possible ways to solve this equation, consider very interesting peculiarities of its solution (which appear in the higher orders of the
series expansion) and propose new superfield actions sharing common features with the action~\p{action1}.

\section{Ketov equation}\label{KEq}
In the analysis of the equation~\p{eq2}, in our opinion, it is preferable to pass to another basis by representing~$\cA_0$ and~$\cbA_0$ as follows:
\be\label{B}
\cA_0= \cW^2 + \bD^4 \cB, \qquad \cbA_0 = \cbW^2 +D^4 \cB.
\ee
In terms of $\cB$ the equation \p{eq2} (together with its conjugated one) could be rewritten as
\be\label{eq3}
\cB=\frac{1}{2} \left( \cW^2 +\bD^4 \cB\right)\left(\cbW^2 +D^4 \cB\right).
\ee
The full action \p{action1} now acquires the form
\be\label{action2}
S = \int d^4 x d^4 \theta \; \cW^2 +\int d^4 x d^4 \bar\theta\;  \cbW^2 +2
\int d^4 x d^8 \theta \; \cB.
\ee
Now, our attention will be devoted to determining the function~$\cB$, in order to construct the action~\p{action2}. As a consequence,  in the solution to eq.~\p{eq3} we can always neglect total derivative terms
\be\label{freedom}
\partial_{\alpha\dot\alpha}\left(\ldots\right),\qquad D^i_\alpha \left(\ldots\right),\qquad
\bD^i_{\dot\alpha} \left(\ldots\right)
\ee
since they do not contribute to the action.

\subsection{Series expansion}
Equation~\p{eq3} can easily be  solved by iterations
\be\label{totalB}
\cB = \sum_{k=0} \cB_k,
\ee
where the indices $k$ count the aggregated power of $\cW$ and~$\cbW$ in the terms $\cB_k$. It is evident that~$\cB$ contains only even powers of~$\cW$ and~$\cbW$. The few first terms read
\be\label{sol1}
\ba{l}
\ds \cB_4 = \frac{1}{2} \cW^2 \cbW^2,\quad \cB_6 = \frac{1}{4} \cW^2 \cbW^2 \left( D^4 \cW^2 + \bD^4 \cbW^2\right) ,\\[0.5em]
\ds \cB_8 = \frac{1}{8} \cW^2 \cbW^2 \left[ \left(D^4 \cW^2\right)^2+ D^4 \cW^2 \bD^4 \cbW^2 + \left( \bD^4 \cbW^2\right)^2 +
\left( D^4 \bD^4 +\bD^4 D^4\right)\left( \cW^2 \cbW^2\right)\right] ,\\[0.5em]
\ds \cB_{10} = \frac{1}{16} \cW^2 \cbW^2 \left[  \left(D^4 \cW^2\right)^3+ \left(D^4 \cW^2\right)^2 \bD^4 \cbW^2 + \rule{0pt}{1.1em}
D^4 \cW^2\left( \bD^4 \cbW^2\right)^2+
 \left( \bD^4 \cbW^2\right)^3 + \right.\\[0.5em]
\ds \phantom{\cB_{10} =} \left( D^4 \bD^4 +\bD^4 D^4\right)\left[ \cW^2 \cbW^2 \left( D^4 \cW^2 +\bD^4 \cbW^2\right) \right]+ \\[0.5em]
\ds \phantom{\cB_{10} =} \left. \left( 2 D^4 \cW^2 + \bD^4 \cbW^2 \right) D^4 \bD^4 \left( \cW^2 \cbW^2\right)+ \left( 2 \bD^4 \cbW^2 +
D^4 \cW^2 \right) \bD^4 D^4 \left( \cW^2 \cbW^2\right)\rule{0pt}{1.1em}\right].
\ea
\ee
Using the freedom~\p{freedom} one may slightly simplify the~$\cB_8$ and~$\cB_{10}$ terms in~\p{sol1} to be
\be\label{sol2}
\ba{l}
\ds \cB_8 = \frac{1}{8} \cW^2 \cbW^2 \left[ \left(D^4 \cW^2\right)^2+ 3 D^4 \cW^2 \bD^4 \cbW^2 +
\left( \bD^4 \cbW^2\right)^2\right], \\
\ds \cB_{10} =\frac{1}{16} \cW^2 \cbW^2\left( D^4 \cW^2+\bD^4 \cbW^2\right) \left[ \left(D^4 \cW^2\right)^2+
D^4 \cW^2 \bD^4 \cbW^2 + \left( \bD^4 \cbW^2\right)^2 + \right. \rule{0pt}{1em}\\
\ds\phantom{\cB_{10} = } \left.  2 \left( D^4 \bD^4 +\bD^4 D^4\right)\left( \cW^2 \cbW^2\right) \rule{0pt}{1em}\right].
\ea
\ee
At this stage one could suppose that the full solution is a function of four arguments
$$D^4 \cW^2,\quad \bD^4 \cbW^2 \quad\mbox{and}\quad D^4 \bD^4 \left(\cW^2 \cbW^2\right), \quad \bD^4 D^4 \left(\cW^2 \cbW^2\right)$$
(in contrast with Ketov's ``solution''~\cite{ketov1} which depends on two arguments only~-- $D^4 \cW^2$ and~$\bD^4 \cbW^2$).
The next order term confirms this
hypothesis
\be\label{B12}\ba{l}
%\bea\label{B12}
\ds\cB_{12}= \frac{1}{32} \cW^2 \cbW^2 \Big \{ \left(D^4 \cW^2\right)^4+ \left( \bD^4 \cbW^2\right)^4
     + \\[0.7em]
\ds\phantom{\cB_{12}= }
    5 \bD^4 \cbW^2\left( D^4 \cW^2+ \bD^4 \cbW^2\right) D^4 \bD^4 \left(\cW^2 \cbW^2\right)+\\[0.7em]
\ds\phantom{\cB_{12}= }
    5 D^4 \cW^2\left( D^4 \cW^2+\bD^4 \cbW^2\right) \bD^4 D^4 \left(\cW^2 \cbW^2\right)+\\[0.7em]
\ds\phantom{\cB_{12}= }
    5 D^4 \bD^4 \left( \cW^2 \cbW^2\right)\; \bD^4 D^4 \left( \cW^2 \cbW^2\right)+\\[0.7em]
\ds\phantom{\cB_{12}= }
    5 D^4 \cW^2 \bD^4 \cbW^2  \left[ \left(D^4 \cW^2\right)^2+ D^4 \cW^2 \bD^4 \cbW^2 + \left( \bD^4 \cbW^2\right)^2\right]
\Big \}.
\ea\ee
%\eea
Alas, the next, 14-th order term $\cB_{14}$ contains new ingredients:
\be\label{B14}
\ba{l}
\ds \cB_{14}=  \frac{1}{64} \cW^2 \cbW^2 \Big \{ \left(D^4 \cW^2\right)^5+  \left( \bD^4 \cbW^2\right)^5 +  \\[0.5em]
\ds \phantom{\cB_{14} = }
 6 D^4 \cW^2 \bD^4 \cbW^2 \left[ \left(D^4 \cW^2\right)^3+ \left(D^4 \cW^2\right)^2 \bD^4 \cbW^2 +
D^4 \cW^2 \left(\bD^4 \cbW^2\right)^2+ \left( \bD^4 \cbW^2\right)^3\right] + \\[0.5em]
\ds \phantom{\cB_{14} = }
 3 \bD^4 \cbW^2 \left( 3\left( D^4 \cW^2\right)^2 +4 D^4 \cW^2 \bD^4 \cbW^2 +
2 \left(\bD^4 \cbW^2\right)^2\right) D^4 \bD^4\left(\cW^2 \cbW^2\right)+ \\[0.5em]
\ds \phantom{\cB_{14} = }
 3 D^4 \cW^2 \left( 3\left( \bD^4 \cbW^2\right)^2 +4 D^4 \cW^2 \bD^4 \cbW^2 +
2 \left(D^4 \cW^2\right)^2\right) \bD^4 D^4\left(\cW^2 \cbW^2\right)+ \\[0.5em]
\ds \phantom{\cB_{14} = }
 2 \bD^4 \cbW^2 \left( D^4 \bD^4\left(\cW^2 \cbW^2\right)\right)^2+
2 D^4 \cW^2 \left( \bD^4 D^4\left(\cW^2 \cbW^2\right)\right)^2+\\[0.5em]
\ds \phantom{\cB_{14} = }
12 \left( D^4 \cW^2+\bD^4 \cbW^2\right) D^4\bD^4\left(\cW^2 \cbW^2\right) \bD^4 D^4 \left(\cW^2 \cbW^2\right)+ \\[0.5em]
\ds \phantom{\cB_{14} = }
3 \left[ \left(\bD^4\cbW^2\right)^2 +2 \bD^4 D^4 \left(\cW^2 \cbW^2\right)\right] D^4 \bD^4
\left(\cW^2 \cbW^2 \bD^4 \cbW^2\right)+\\[0.5em]
\ds \phantom{\cB_{14} = }
 3 \bD^4 D^4 \left(\cW^2 \cbW^2\right) D^4 \bD^4 \left(\cW^2 \cbW^2 D^4 \cW^2\right)+\\[0.5em]
 \ds \phantom{\cB_{14} = }
3 \left[ \left(D^4 \cW^2\right)^2 +2 D^4 \bD^4 \left(\cW^2 \cbW^2\right)\right] \bD^4 D^4 \left(\cW^2 \cbW^2
D^4 \cW^2\right)+\\[0.5em]
\ds \phantom{\cB_{14} = }
 3 D^4 \bD^4 \left(\cW^2 \cbW^2\right) \bD^4 D^4 \left(\cW^2 \cbW^2 \bD^4 \cbW^2\right)\Big \}.
\ea
\ee
The last four lines contain new terms which cannot be represented in terms of the four independent variables we already
introduced.
When constructing the next orders, the situation becomes worse:  in the solution further and further new objects appear.
%and this story will never end up.
Thus, the first lesson we learn from considering higher order terms is that the full solution cannot be represented as some function depending on {\it a finite number} of its arguments.

\subsection{Linear approximation}
In order to understand the mechanism of generating new structures in the action one could further simplify equation~\p{eq3} discarding the nonlinear term proportional to $\cB^2$:
\be\label{eq4}
\cB=\frac{1}{2} \cW^2 \cbW^2 +\frac{1}{2}\left( \cW^2 D^4 \cB + \cbW^2 \bD^4 \cB\right).
\ee
This linear part in $\cB$ equation can immediately be solved as
\be\label{solB1}
\cB= \frac{1}{2}\; Y\left[\, \strut \cW^2\, \cbW^2\,\right]
%\equiv \; \frac{\frac{1}{2}}{1- \left(\frac{1}{2}X +\frac{1}{2}\bX\right)}\;\cW^2 \cbW^2,
\ee
in terms of the pseudo-differential operator~$Y$
%where we introduced the operators $Y$ and $X, \bX$ as
\be
Y\equiv  \frac{1}{1-\frac{1}{2}\left(X +\bX\,\right)}, \qquad
X=\cW^2 D^4,\quad \bX=\cbW^2 \bD^4.
\ee
The action of the operator $Y$ on an arbitrary superfield $\cF$ is understood as a formal series of operators
$X,\bX$
\be
Y \left[\cF\right] \equiv \cF+\frac{1}{2}\left(X +\bX\right)\left[\cF\right]
    +\frac{1}{2^2}\left(X + \bX\,\right)\left(X + \bX\,\right)\left[ \cF\right]+\ldots
\ee
It is instructive to understand what terms appear in the action \p{action2} with $\cB$ given by \p{solB1}:
\be\label{sol_free}\ba{l}
\ds \cB_4 = \frac{1}{2} \cW^2\cbW^2, \qquad \cB_6  = \frac{1}{4} \cW^2 \cbW^2 \left( D^4 \cW^2 +\bD^4 \cbW^2\right), \\[0.5em]
\ds \cB_8 = \frac{1}{8} \cW^2\cbW^2\left[ \left( D^4 \cW^2\right)^2 +2 D^4 \cW^2\; \bD^4 \cbW^2 +
\left( \bD^4 \cbW^2\right)^2\right], \\
\ds \cB_{10} = \frac{1}{16}\Big \{ \cW^2\cbW^2 \left( D^4 \cW^2 +\bD^4 \cbW^2\right)\left[ \left( D^4 \cW^2\right)^2 +
D^4 \cW^2\; \bD^4 \cbW^2 +\left( \bD^4 \cbW^2\right)^2\right]+ \\[0.5em]
\ds \phantom{\cB_{10} = }
\cW^2 \cbW^2 \left( \bD^4 \cbW^2 D^4 \bD^4 \left(\cW^2 \cbW^2\right)+
D^4 \cW^2 \bD^4 D^4 \left(\cW^2 \cbW^2\right)\right)\Big \}.
\ea\ee
Thus we see that the needed new terms are indeed generated by the action of operators~$X$ and~$\bX$ on~$\cB_4$.
Obviously, this solution is not a solution to Ketov equation, it is just to illustrate the idea of using operators to reproduce new structures in the action.

One should stress that, despite the linearity  of equation~\p{eq4}, the resulting action is highly nonlinear and non-trivial, due to the nonlinearity of the Lagrangian density~$\cB$ in terms of~$\cW$.

\subsection{``Classical'' limit}
From the above considered linear approximation we see the necessity of introducing the complex operator~$X$. This passage resembles that in quantum mechanics, when passing from functions to operators. We try to make some use of this analogy. With the help of this operator, eq.~\p{eq3} acquires the following form:
\be\label{eq3a}
\cW^2 \cbW^2\; \cB =\frac{1}{2} \left( \cW^2 \cbW^2\right)^2+ \frac{1}{2}\cW^2 \cbW^2\left (X+\bX\right) \cB+
\frac{1}{2}\left(X\; \cB\right) \left( \bX\; \cB\right).
\ee
Now one may consider a ``classical'' version of this equation replacing the operators $X,\bX$ by their ``classical'' analogs
$$%\be\label{subst}
X \rightarrow x \equiv \left( D^4 \cW^2\right)
,
\qquad \bX \rightarrow \bx \equiv \left( \bD^4 \cbW^2\right)
,
$$%\ee
defined in~\p{var0}. This replacement transforms equation~\p{eq3a} into an algebraic one, whose solution can be immediately found
\be\label{solK}
\cB = \Upsilon(x,\bx)\; \cW^2 \cbW^2 =
    \frac{ 2-\left( x+\bx\right) - \sqrt{ \left( 2 -\left(x+\bx\right)\right)^2 - 4\, x\,\bx}}{ 2\, x\, \bx}\; \cW^2\cbW^2.
\ee
Now it becomes obvious that the action~\p{action2} with the expression~$\cB$ given by eq.~\p{solK} is exactly the action proposed by S.~Ketov \cite{ketov1}.
Clearly, it is not a solution to equation~\p{eq3a}, but rather a solution to its ``classical'' limit, obtained by the {\it unjustified} replacement of the operator~$X$ by its ``classical'' expression~$x$.

One should note that the exact solution of equation~\p{eq3} goes into~\p{solK} in the ``classical'' limit. % \p{subst}.
Therefore, the expression~\p{solK} for~$\cB$, no doubts, contains important information about  the exact solution.

\subsection{First approximation}
Now one can try to guess an operational variant of the solution considering the following Ansatz
\be\label{qu1}
\cB_q = \Upsilon(X,\bX) \cW^2 \cbW^2.
\ee
This Lagrangian density has a proper ``classical limit''~\p{solK} and, at the same time, it contains a lot of new terms
which cannot be reproduced by the solution~\p{solK}. One can check that the exact Lagrangian density~\p{sol1} is correctly
reproduced, up to the 10-th order in the field strengths $\cW, \cbW$.

In order to demonstrate how the terms present in eqs.~\p{sol2} can be generated by the Ansatz~\p{qu1}, let us consider, for example, the following term:
$$%\be\label{test}
\cW^2 \cbW^2 \left( D^4 \cW^2\right)^2 \bD^4 \cbW^2
$$%\ee
By power counting considerations, one deduces that the only possibility to reproduce this term  is to act twice by the operator~$X$ and once by~$\bX$ :
$$%\be\label{test2}
X\, X\, \bX \,\left[ \cW^2 \cbW^2\right] \equiv \cW^2 D^4 \left[ \cW^2 D^4 \left[ \cbW^2 \bD^4 \left[ \cW^2 \cbW^2\right]\right]\right] =\cW^2 \cbW^2 \left( D^4 \cW^2\right) D^4 \left[ \cW^2 \bD^4 \cbW^2\right].
$$
Modulus total derivative terms described in~\p{freedom}, this expression reads
$$
D^4 \left[ \cW^2 \cbW^2 \left( D^4 \cW^2\right)\right]  \cW^2 \bD^4 \cbW^2 = \cW^2 \cbW^2 \left( D^4 \cW^2\right)^2 \bD^4 \cbW^2.
$$
One sees that freedom~\p{freedom} plays an important role in our construction.

\subsection{Next approximations}
Alas, the promising Ansatz~\p{qu1} reproduces only a part (but still an infinite one) of the needed terms. The differences between~\p{qu1} and the exact expressions~\p{B12} appears already in the next, 12-th order. At this order, a correction has the following form:
\be\label{test4}
\cB_{12} = \left( \cB_q \right)_{12} - \frac{5}{192} \left( X\; X_3 \; X_3 \; \bX + \bX\; X_3\; X_3 \; X\right)\left[
\cW^2 \cbW^2\right] ,
\ee
where a new operator~$X_3$ is defined as
\be\label{x3}
X_3\left[ \cF \right] = \cW^2 \left( D^4 \cF\right)  + \cbW^2  \left( \bD^4 \cF\right)
    - \left(D^4 \cW^2\right) \cF - \left( \bD^4 \cbW^2\right) \cF.
\ee
This operator vanishes in the classical limit and thus it is indeed a purely ``quantum'' one.
In the next orders new corrections appear with the same structure as in~\p{test4}. Namely, these new terms contain the correct number of operators~$X$ and~$\bX$ and only two operators $X_3$. Moreover, these new additional terms cannot be represented as a function of generators (in contrast to the expression~$\cB_q$ in~\p{qu1}). Thus, starting from the 12-th order the non-commutativity of the operators~$X$,~$\bX$ and~$X_3$ appears to be crucial.

It is funny enough that with these three operators one could successfully construct all terms in the exact solution, up to the
18-th order included. In principle, one should even suggest the form of new corrections needed in each order.
Unfortunately, in the 20-th order (the highest which we were able to check) a new ``quantum'' structure is needed. It is not even an
operator~-- it is just a new variable
\be\label{finv}
x_4 = \left( \left[ D^4, \bD^4 \right] \cW^2 \cbW^2 \right).
\ee
which we have to introduce to reproduce a new structure which shows up only in the 20-th order. Again, it is a pure quantum
variable: it becomes zero in the classical limit, but it is absolutely needed. The necessity of this new variable
makes the whole analysis quite cumbersome and unpredictable, because we cannot forbid the appearance of this variable
in the previous lower orders to produce the structures already generated through the action of operators~$X, \bX$ and~$X_3$.

\section{Conclusion}
In this paper we investigated in details the structure of the exact solution of Ketov equation, which contains
important information about~$N=2$ supersymmetric Born-Infeld theory. Analyzing higher orders terms in the
iterative solution to this equation we have found that, on each level, new structures arise. Thus, it seems impossible to
write the exact solution as a function depending on {\it a finite number} of its arguments. We proposed to introduce
differential operators which could, in principle, generate new structures for the Lagrangian density.
Using these differential operators we were able to reproduce the corresponding Lagrangian density, up to the 18-th order
in the series expansion over~$N=2$ field strengths~$\cW$. Unfortunately, the highest order that we managed to deal with (the 20-th order)
calls for new structures which cannot be generated by the action of the generators $X$ and~$X_3$.

We proposed two new superfield actions which share common features with the ~$N=2$ supersymmetric BI action: the latter is their particular limit. These actions contain an infinite number of terms which come from the action of a pseudo-differential operator on the superfields. The proposed action~\p{qu1} turns into the Ketov's one in the so-called ``classical'' limit. So, probably, it contains the bosonic Born-Infeld action.

These new actions catch some important properties of the exact solution of the Ketov's equation. Moreover, we have at hands  some nice, closed form for these actions. Unfortunately, they contain an infinite number of terms. So,  it is not clear how to analyze these actions, even in the bosonic limit. Clearly, the same is true also for the exact action of~$N=2$ supersymmetric BI theory. Thus, we have to conclude that the standard framework where the basic superfields obey linear chirality conditions is not very useful. We believe that the nonlinear realization approach with nonlinear chirality conditions imposed on the~$N=2$ field strengths~\cite{BIK1,BIK2,BIK3} will be more adequate, in order to construct the full action of supersymmetric BI theory.

\section*{Acknowledgements}
S.Krivonos and A.Sutulin are grateful to the Laboratori Nazionali di Frascati for warm hospitality. This work was partially supported by RFBR grants~11-02-01335,~12-02-00517 and~11-02-90445-Ukr as well as by the ERC Advanced Grant no. 226455 \textit{``Supersymmetry, Quantum Gravity and Gauge Fields''}~(\textit{SUPER\-FIELDS}).

\end{document}